\documentclass[12pt,letterpaper]{article}
\pdfoutput=1
\usepackage{fixltx2e} \usepackage{textcomp} \usepackage{fullpage} 
\usepackage{verbatim} \usepackage[english]{babel} \usepackage{pifont}
\usepackage{xcolor} \usepackage{setspace} \usepackage{lscape} 
\usepackage{indentfirst} \usepackage[normalem]{ulem}
\usepackage{booktabs} 
\usepackage{float}	\usepackage{latexsym}   
\usepackage{url} 
\usepackage{hyperref} \usepackage{epsfig} \usepackage{graphicx}
\usepackage{amssymb} \usepackage{amsmath} \usepackage{amsfonts} \usepackage{amsthm} \usepackage{amstext}
\usepackage{bm} \usepackage{array} \usepackage{mhchem}
\usepackage{ifthen} \usepackage{caption}
\usepackage[applemac]{inputenc}
\usepackage{enumerate}

\raggedright
\renewcommand{\section}[1]{%
\bigskip
\begin{center}
\begin{Large}
\normalfont\scshape #1
\medskip
\end{Large}
\end{center}}

\renewcommand{\subsection}[1]{%
\bigskip
\begin{center}
\begin{large}
\normalfont\itshape #1
\end{large}
\end{center}}

\renewcommand{\subsubsection}[1]{%
\vspace{2ex}
\noindent
\textit{#1.}---}

\renewcommand{\tableofcontents}{}

\usepackage{tikz}        
\usepackage{xcolor}

\newcommand {\Real} {\mathbb{R}}	
\newcommand{\veps}{\varepsilon}

\newcommand {\bydef}{\,\raise.07485ex\hbox{:}\kern-.025em\hbox{=}\,}
\newcommand {\Ac}  {\mathcal{A}}
\newcommand {\Bc}  {\mathcal{B}}
\newcommand {\Cc}  {\mathcal{C}}

\newcommand {\Ec}  {\mathcal{E}}

\newcommand {\Mc}  {\mathcal{M}}
\newcommand {\Nc}  {\mathcal{N}}

\newcommand {\Tc}  {\mathcal{T}}
\newcommand {\Uc}  {\mathcal{U}}
\newcommand {\Vc}  {\mathcal{V}}

\newcommand {\eb} {\mathbf{e}}

\newcommand {\Bo} {\mathbb{B}}

\begin{document}
\bigskip

\bigskip
\medskip
\begin{center}

\noindent{\Large \bf Comparing Trajectories on the Size and Shape Space}
\bigskip



\noindent {\normalsize \sc 
 Valerio Varano$^1$, Stefano Gabriele$^1$, Luciano Teresi$^2$, Ian Dryden$^3$,  Paolo E. Puddu$^4$, Concetta Torromeo$^4$ \\ and Paolo Piras$^{4,5,6}$}\\
\noindent {\small \it 
$^1$Dept. Architecture, Roma Tre University, Roma, Italy;\\
$^2$Dept. Mathematics \& Physics, Roma Tre University, Roma, Italy;\\
$^3$School of Mathematical Sciences, University of Nottingham, Nottingham, UK;\\
$^4$Dipartimento di Scienze Cardiovascolari, Respiratorie, Nefrologiche, Anestesiologiche e Geriatriche, Sapienza, Rome, Italy;\\
$^5$Dept. Sciences, Roma Tre University, Roma, Italy;\\
$^6$Center for Evolutionary Ecology, Isernia, Italy.}\\
\end{center}
\medskip
\noindent{\bf Corresponding author:} Valerio Varano 
LaMS, Roma Tre University, 
via della Madonna dei Monti, 40 - Roma, Italy, 00184; 
E-mail: valerio.varano@uniroma3.it\\


\vspace{1in}
%
\subsubsection{Abstract}
%
In this paper we show that trajectory shape analysis should be performed only after obtaining a proper representation before applying ordination methods. In fact, studying the shape of a trajectory means studying how the deformation changes along each path irrespectively of the actual shape to which these deformations apply. The independence of the deformation from the shape to which it is applied is critical: it implies that any shape variation between individuals at the beginning of each trajectories must be completely filtered out. A Parallel Transport, that can be based on various connection types, is necessary to perform such kind of shape data centering. The Levi Civita connection  can also be used to transport a deformation. We demonstrate that this procedure does not preserve deformation even in the affine case. We propose a novel procedure called ”Direct Transport” able to perfectly transport deformation in the affine case and to better approximate non affine deformation in comparison to existing tools. 
%
\par\noindent (Keywords: Riemannian Manifold, Deformation cycle, Parallel Transport, Trajectory Analysis, Equipollence, Thin Plate Spline, Inter-individual difference, Geometric Morphometrics )\\
\vspace{1.5in}
%
%

\section{Introduction}
Since \cite{Thompson1917} the study of shape change has become a central topic in the science of form. Modern morphometrics, i.e. Geometric Morphometrics (GM),  begins with the seminal contributions of \cite {Kendall1977,Kendall1984}. He proposed a criterion to eliminate all non shape informed differences to look for dissimilarity between two or more shapes. In the case of shapes defined by homologous landmarks, the non shape informed  attributes are size, translation and rotation. The first two are measured by Centroid Size (CS) and Centroid displacement respectively and are eliminated by scaling all shapes to a common CS (usually 1) and by centering their Centroids. Several criteria exist for eliminating rotation, via the alignment of homologous landmarks and the choice of a metric that needs to be minimized during the superimposition of all configurations. This alignment often happens via an iterative algorithm (Bookstein, 1987; Rohlf and Slice, 1990). 
One of the most frequently used criteria for alignment is the Procrustes Distance metric used in the Generalized Procrustes Analysis (GPA). \cite{Kendall1984} shows how this metric is a geodesic distance on a Riemannian differentiable multidimensional manifold of $m\times (k-1)-1-m(m-1)/2$ dimensions, were $m$ is the number of dimensions and $k$ the number of landmarks. This manifold is named Shape Space. The pole of the shape space is usually set on the average of all configurations, often called  the “consensus”. When the variation around the consensus is small, the geodesic Procrustes Distance is approximated by its projection (usually orthogonal) on a tangent plane to the consensus. The aligned coordinates are then often subjected to ordination methods, such as Principal Component Analysis (PCA) for further analysis . \cite{Rohlf2000} showed that methods other than those based on Procrustes Distances strongly constrain the possible results obtained by ordination analyses and can give misleading results when used in studies of growth and evolutionary trajectories. 

In this paper we focus just on the study of \emph{morphological trajectories} $\Tc$, defined as an 
ordered (upon a given criterion, i.e. time, size or other) sequence of shapes. 
Given a set of different morphological trajectories $\Tc_j$, 
it is possible to study the structure of the trajectories themselves. 
This is particularly desirable because this investigation could shed light on the peculiarities of the deformation
processes, that may be independent from the shape differences among corresponding shapes belonging to different
trajectories. 
As an example, denoting with $s$ a shape, two morphological trajectories $\Tc_1=(s_1, ..., s_n)$ 
and $\Tc_2=(\bar s_1, ..., \bar s_n)$ 
could be manifestations of a same deformative process, irrespective of the fact that $s_1\ne \bar s_1$.

A rationale for this Phenotypic Trajectory Analysis (PTA) has been proposed by \cite{Adams2009} and \cite{Collyer2013}. They suggested that a trajectory has specific attributes such as shape, size and orientation that can be studied via shape analysis. This is possible if specific steps of each trajectory can be recognized as homologous. This homology criterion is crucial and can be time-based or physiology-based in biological applications (\cite{Piras2014}). Once assessed the possibility to use some (or all) steps as “landmarks” that can be considered homologous among different trajectories, the study of their shapes can proceed via GPA. 

In this paper we show that trajectory shape analysis should be performed only after obtaining a proper representation before applying ordination methods. In fact, studying the shape of a trajectory means studying how the deformation changes along each path irrespectively of the actual shape to which these deformations apply. The independence of the deformation from the shape to which it is applied is critical: it implies that any shape variation between individuals at the beginning of each trajectories must be completely filtered out. Often, in statistics, inter-group differences are eliminated by applying a group-mean centering, optionally followed by the Grand Mean addition. A problem arises if the data are shape data. 
Very frequently the Levi Civita connection on the shape space is used to compute the geodesics between two shapes (\cite{Kume2007}, \cite{Le2003}). Sometimes it is also used to transport a deformation along this geodesic, in order to apply a deformation of the first shape to the second shape (\cite{Le2000}, \cite{Huckemann2010}). 
Formally this procedure could be applied in order to center data in the Shape Space, but it is revealed to be inadequate in some cases because it does not conserve the physical meaning of the deformation during the path. Many efforts have been done in recent years in order to unify shape metrics with deformation metrics, especially in the field of image matching and pattern recognition (\cite{Peter2009}, \cite{Miller2001}). In particular new metrics have been proposed together with the corresponding induced LC connections. In the present article we show that a special procedure that we call “Direct Transport” is useful in order to compare different morphological trajectories by performing a data centering which maintains the deformations. We also show that the DT procedure can be built, for the case of affine deformations, on a formalized new connection on the size-and-shape space, different from the LC one. We used simulated datasets and an a priori known set of affine and non affine deformation parameters in order to build properly pre-processed morphological trajectories to be used in standard ordination methods such as PCA. In addition we illustrate the methodology with an application in cardiology, which motivate the work.    
\section{The geometrical structure of the shape space}\label{sec_geom}
%
%
A body $\Bc$ is an open subset of the $m$-dimensional Euclidean ambient space $\Ec_m$; its shape is sampled through
the position $x\in \Ec_m$ of $k$ points, called \emph{landmarks}. 
A \emph{configuration} of $\Bc$ is defined as the ordered $k$-ple of landmarks, which can be represented as a
$k\times m$ matrix $X=(x_1,\ldots,x_k)^T$; we denote with $\Cc^k_m$ 
the \emph{configuration space}, that is, the set of all possible configurations. 

The \emph{Shape Space} $\Sigma^k_m$ can be defined as the quotient of  $\Cc^k_m$ under the action of 
the group $\mathcal{S}(m)$ of the Euclidean similarity transformations in $\Ec_m$. $\mathcal{S}(m)$ can be decomposed in three subgroups: translations 
$\Vc\Ec_m$; rotations $SO_m$; homothety or dilatation $\mathcal{H}_m$. 
The shape space can be conveniently generated by removing similarity transformations one by one; 
this construction can be done in at least two alternative ways, yielding as consequence to the definition of two different intermediate spaces (\cite{Dryden1998}). 
\begin{itemize}
\item{}
The first way is that formulated by Kendall: first remove translations and size to obtain the \emph{pre-shape space} 
$S^k_m\equiv S^{(k-1)m-1}$; then removes rotations to get the shape space as $\Sigma^k_m=S^k_m/SO_m$. 
\item{}
The second way removes at first rigid transformations--translations and rotations--thus 
obtaining the \emph{size-and-shape space} $S\Sigma^k_m$; then removes size to obtain the shape space 
as $\Sigma^k_m=S\Sigma^k_m/\mathcal{H}_m$. 
\end{itemize}
In this paper translations are filtered out by using centered configurations $X_C$ defined as follows:
\begin{equation}\label{centered}
X_C=C X\,,
\end{equation}  
with $C=I_k-\frac{1}{k}1_k 1_k^T$, where $I_k$ is the $k\times k$ identity matrix, and $1_k$ is a $k\times 1$ 
column of ones.
The centered pre-shape $Z_C \in \,S^k_m$ of a configuration $X$ is defined as:
\begin{equation}\label{pre-shape}
Z_C=\frac{X_C}{||X_C||}\,,
\end{equation}  
where $||X_C||=\sqrt{\text{trace}(X_C^T\,X_C)}$ is the centroid size of $X_C$.
Finally, the shape of a configuration $X$ is the equivalence class $[X]\in\,\Sigma^k_m$ 
defined by:
\begin{equation}\label{shape}
[X]=\{ Z_C\,Q:Q\in SO_m \}\,.
\end{equation}
On the other hand, the size-and-shape of a configuration $X$ is the equivalence class $[X]_S\in\,S\Sigma^k_m$ 
represented by:
\begin{equation}\label{size-and-shape}
[X]_S=\{ X_C\,Q:Q\in SO_m \}\,.
\end{equation}
From (\ref{pre-shape}), (\ref{size-and-shape}) it follows $[X]=[X]_S\,/\,||X_C||$. 

Let us note that the procedure described above defines implicitly an atlas for the shape space,
which inherits a manifold structure. Actually, the Kendall shape space has a richer geometric structure, 
being endowed with a Riemannian structure, a metric, and a connection on the tangent bundle; 
these structures are defined by a metric tensor ($g$) on the tangent bundle, a distance ($d$) on the manifold and a covariant derivative ($\nabla$) on the tangent bundle, respectively; 
Table \ref{geom_struc} summarizes this richness in geometrical structure.

It is important to stress that these choices are independent of each other, and the richness of the resulting 
geometric structure is overshadowed by both the elegance of the Kendall's construction, and by 
the tacit identification $\Cc^k_m\equiv\Ec_{k\,m}$ of the configuration space $\Cc^k_m$ 
with the $k\times m$-Euclidean one $\Ec_{k\,m}$; 
in particular, it is assumed that this identification holds for each level of the geometrical structure (each row of the Table \ref{geom_struc}).  The meaning and the consequences of this assumption will be discussed in the following; 
here we take it for granted.
\begin{table}[h]
\begin{center}
\begin{tabular}{|l|c|c|c|c|}
\hline
          						  & $\Cc^k_m$   	  &$S^k_m$	         & $S\Sigma^k_m$ 	&$\Sigma^k_m$  \\[2mm]
\hline
Manifold (Atlas)           		  & $X$    		  & $Z$    		     & $[X]_S$			&$[X]$		   \\[2mm]
Riemannian Space (Metric tensor)	  & $g$     	      & $g_Z$		     & $g_S$			    &$g_\Sigma$	   \\[2mm]
Metric Space   (Distance)      	  & $d$     		  & $\rho_Z$         & $d_S$		     	&$\rho$		   \\[2mm]
Connection (Covariant derivative) & $\nabla^{LC}$ &  $\nabla_Z^{LC}$	 & $\nabla_S^{LC}$	& $\nabla_\Sigma^{LC}$	\\[1mm]
\hline
\end{tabular}
\caption{The whole structure of the shape space}\label{geom_struc}
\end{center}
\end{table}
Once accepted that the entire geometrical structure of $\Ec_{k\,m}$ is inherited by $\Cc^k_m$,
one observes that the regular part of the shape space $\Sigma^k_m$ is built by a sequence of Riemannian isometric maps: 
an ortogonal projection (immersion), followed by a quotient (submersion), or vice versa, as illustrated in 
(\ref{sequences}).  
\begin{eqnarray}\label{sequences}
 \nonumber &\Cc^{km}\equiv \Ec^k_m \xrightarrow[\text{ortogonal projection}]{\text{isometric immersion}}& S^k_m \xrightarrow[\text{quotient}]{\text{isometric submersion}} 	\Sigma^k_m\,,\\[3mm]
&\Cc^{km}\equiv \Ec^k_m \xrightarrow[\text{quotient}]{\text{isometric submersion}}& S\Sigma^k_m \xrightarrow[\text{ortogonal projection}]{\text{isometric immersion}} 		\Sigma^k_m\,.
 \end{eqnarray} 
This sequence induces isometrically all the geometric structure from the configuration space $\Cc^{km}$ 
to the shape space $\Sigma^k_m$; 
all the details can be found in (\cite{Kendall1999}, \cite{Le2003}, \cite{Huckemann2010leaf}).

Here, it is useful to recall that in the Euclidean space $\Ec_{k\,m}$, the tangent spaces at any point
can be identified, thus yielding a global vector space, the translation one $\Vc\Ec_{k\,m}$, thus, to each pair of
points $(X_1,\,X_2)$ there correspond a vector $V=X_2-X_1 \in\,\Vc\Ec_{k\,m}$;  
the Euclidean metric tensor $g(V_1,V_2)=\text{trace}(V_1^T V_2)$ is then naturally used to define
an Euclidean distance:
\begin{equation}
 d(X_1, X_2)=||X_2-X_1||=\sqrt{g(X_2-X_1,X_2-X_1)}\,.
\end{equation} 
Without entering into details (which will be given in the next section),
here we complete the picture of the Euclidean space structure by recalling that the connection on 
$\Ec_{k\,m}$ is the so called Levi Civita (LC) connection, defined through the 
Levi Civita covariant derivative $\nabla^{LC}$.
The induced distance on the size-and-shape space $S\Sigma^k_m$ results to be:
\begin{equation}
d_S(X_1,X_2)=\inf_{Q\in SO(m)} ||X_2-X_1Q||\,.
\end{equation}
Finally, the projection on the hypersphere gives the Procrustes distance: 
\begin{equation}
\rho(X_1,X_2)=\arccos\left(\frac{d_S(X_1,X_2)^2-S_1^2-S_2^2}{2 S_1S_2}\right) \,,
\end{equation}
where $S_1=||C X_1||$, $S_2=||C X_2||$ are the centroid sizes of the configurations $X_1$ and $X_2$.
%
\subsection{Morphological trajectories}
%
Given an ordered sequence of bodies $\Bo=\{\Bc_1,\ldots,\Bc_n)$, 
and denoted with $X_i$ the configuration of $\Bc_i$, we can consider the associated sequences:
\begin{equation}
\begin{array}{ll} 
\textrm{\emph{Trajectory} of } \Bo\,:  & \Tc=(X_1,\ldots,X_n)\,;\\[3mm]
\textrm{\emph{Morphological trajectory} of } \Bo\,: & \Mc\Tc=([X_1],\ldots,[X_n]) \,. 
\end{array}
\end{equation}
Let us note that a trajectory $\Bo$ can be considered, as a whole, to represent a body sampled with 
$k\times n$ ordered landmarks: thus, the shape of a trajectory is worth investigating.

A basic notion that will be crucial in the following is that of deformation, a
smooth, bijective, and interpolant map $\Phi: \Ec_m \to \Ec_m$.
Given a pair of configurations $X, Y\in\, \Cc^k_m$, we say that $Y$ is a deformation of $X$ if
\begin{equation}\label{diffeo}
y_i=\Phi(x_i)=(\Phi_1(x_i), ..., \Phi_m(x_i))\,,
\quad \forall\,
x_i\in X\,, y_i\in Y\,.
\end{equation}
Here X is the source and Y is the target.
Note that the deformation is from the whole space $\Ec_m$ to $\Ec_m$, rather than just from a set of landmarks. 
We can say that a deformation $\Phi$ is a diffeomorphism from $\Ec_m$ to itself. 
Furthermore, note that the deformation is a notion pertaining to the configuration space rather than to the shape space. 
A family of deformations $\Phi_t: \Ec_m \to \Ec_m$, smoothly parametrised by a scalar $t$, is called \emph{motion}.
Given a motion, we can generate a sequence of bodies by deforming a given body $\Bc$; we define:
\begin{equation}
\begin{array}{ll} 
\textrm{Motion of $\Bc$ along $\Phi_t$:} & \Bo_\Phi(\Bc)=\{\Phi_{t_1}(\Bc),\ldots,\Phi_{t_n}(\Bc)\} \, ;  \\[3mm]
\textrm{Trajectory of } \Bo_\Phi(\Bc):   & \Tc_\Phi(\Bc)=(X_{t_1},\ldots,X_{t_n})\,, 
								\textrm{with $X_{t_i}$ configuration of $\Phi_{t_i}(\Bc)$}\,.
\end{array}
\end{equation}
We shall tackle two main examples:
\begin{enumerate}
\item{} Different motions of the same body: given different motions $\Phi^j_t$, and a single body $\Bc$,
we can generate many different trajectories:
$$
\Tc_{\Phi^j}=(X^j_{t_i})\,,  
\quad\textrm{with $X^j_{t_i}$ configuration of $\Phi^j_{t_i}(\Bc)$};
$$
\item{} Same motion of different bodies: given a motion $\Phi_t$ and different bodies $\Bc_j$,
we generate many different trajectories:
$$
\Tc_\Phi=(X^j_{t_i})\,,  
\quad\textrm{with $X^j_{t_i}$ configuration of $\Phi_{t_i}(\Bc_j)$};
$$ 
\end{enumerate}
Please, note that the apex `j' in $X^j_{t_i}$ can refer both to a specific motion $\Phi^j_t$, as in item 1,
or to a body $\Bc_j$, as in item two. 

Our goal is the development of a procedure to compare morphological trajectories, able to discriminate between intra- and inter- subject variations. 
If the displacements between the shapes of a morphological trajectory are small enough, 
they can be considered as vectors belonging to a same tangent space of $\Sigma_m^k$; when
such is the case, morphological variations can be efficiently assessed by ordination analyses as the PCA performed on the covariance matrix.
The problem arises when two or more morphological trajectories span different and distant neighborhoods 
of the shape space.
%
\section{Transporting Configurations}
%
%
%
%
%
When morphological trajectories lie in distant neighbourhoods of the shape space, to compare 
trajectories it is necessary to transport displacement vectors from a tangent space to another. 
In differential geometry such an operation is called parallel transport, and is based on a connection on the tangent bundle of the manifold. In the case of a Riemannian manifold, it is usual to require that the parallel transport is compatible 
with the metric in the sense that it is an isometry with respect to the given Riemannian metric. 
Such type of connection is named Riemannian connection (see \cite{Klingenberg1982}). 
	
To be more precise, given a manifold $\Mc$,  and an interval $[a,b]\in \Real$, 
the  parallel transport  along a curve $\gamma: [a,b]\to\Mc$  is a path-dependent isomorphism between the tangent spaces
$\Tc\Mc|_{\gamma(t)}$ along the curve:
\begin{equation}\label{tau}
\tau_t: \Tc\Mc|_{\gamma(a)}\to\Tc\Mc|_{\gamma(t)}\hspace{2cm}V_a\mapsto V_t
\end{equation}
In practice, $\tau_t$ must be a one to one linear transformation, 
i.e. must hold $(V+W)_t=V_t+W_t$, $(\lambda V)_t = \lambda V_t$ for $\lambda\in\Real$, 
and $\tau_t^{-1}$ must exist for each $t\in[a,b]$.
Usually $\tau_t$ is defined by a covariant derivative $\nabla$ along the curve. 
A covariant derivative--or a \emph{connection}--on $\Mc$ is encoded in the Christoffel symbols $\Gamma_{ij}^k$:
\begin{equation}
\nabla_{\eb_i} \eb_j= \Gamma_{ij}^k\eb_k\,,
\end{equation}
where $\eb_j$ are the vector fields basis of the tangent bundle. 
A vector field $V$ is said to be parallel along the curve $\gamma$ if holds:
\begin{equation}\label{ODE}
\nabla_{\dot\gamma} V= 0
\end{equation}
Once represented a vector as $V=V^i\eb_i$: 
\begin{equation}
\dot V^k(t)+V^j(t)\,\dot\gamma^i(t)\, \Gamma_{ij}^k\,(\gamma(t))=0
\end{equation}
are linear differential equations with unique solutions $V^j(t)$. 
It is well known that there is locally a unique solution $V^j$ along $\gamma$ for a given initial value.

As shown in \cite{Spivak1999}, nevertheless the parallel transport is usually defined in terms of 
the connection $\nabla$, 
one can also reverse the process: assume a parallel transport $\tau$, and define the connection by a limit: 
\begin{equation}\label{tau_to_nabla}
\nabla_{V_p}U =\lim_{h \to 0}\frac{\tau_h^{-1}U_{\gamma(h)}-U_{\gamma(0)}}{h}
\end{equation}
A connection is called compatible with a metric $g$ if the parallel transport is an isometry, 
i.e. $g_{\gamma(a)}(V_a,W_a)=g_{\gamma(t)}(V_t,W_t)$ for each pair of vector fields $V,W$ and for each $t$.
The \emph{torsion} $T$ of the connection $\nabla$ is a tensor defined as:
\begin{equation}
T(V,W)=\nabla_V W - \nabla_W V -\left[ V,W\right]\,.
\end{equation}	
A connection is called symmetric when $T(V,W)=0$, $\forall V,W$; this yields $\Gamma_{ij}^k=\Gamma_{ji}^k$.
A fundamental lemma of the Riemannian Geometry states that there is a unique symmetric connection 
compatible with the metric $g$, named Levi Civita (LC).
Usually the proof of this lemma is given by obtaining the expression of the 
Christoffel symbols in terms of those of the metric tensor: 
\begin{equation}\label{th_egregium}
\Gamma_{ijq} = \frac{1}{2}\,\Bigl(\partial_j\,g_{iq} + \partial_i\,g_{jq} - \partial_q\,g_{ij}\Bigr)\,.
\end{equation}
with $\Gamma^p_{ij} = g^{pq}\,\Gamma_{ijq}$ 
	\begin{figure}
	\begin{center}
	\includegraphics[scale=0.6]{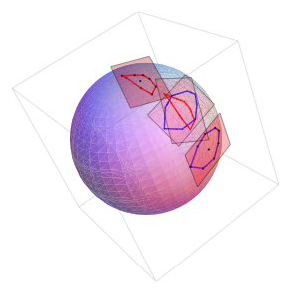}
	\end{center}
	\caption{Data centering via parallel transport.}
	 \label{PT}
	\end{figure}
%
%
%
%
%
%
The uniqueness of the LC connection allows us to transfer easily a connection from a Riemannian space to another 
via isometric maps. 

Since the work of \cite{Kendall1984}, the LC connection on the Shape Space $\Sigma_m^k$ 
has been widely studied. 
As outlined in the previous Section the regular part of the Shape Space can be defined by means 
of a sequence of Riemannian immersions and submersions starting from the Configuration Space $\Cc_m^k$, 
so  that LC connection on Shape Space can be isometrically inherited from that on the Configuration Space. 
In particular, the LC connection on the Pre-shape Space $S\Sigma_m^k$ can be simply derived by a projection, 
while the LC connection on Shape Space can be derived from the previous one by using the O'Neill theorem on submersions (\cite{Le2003}).
 
When $m=2$ (2D bodies), the parallel transport has an explicit representation, 
while for $m=3$ (3D bodies), the parallel transport can be performed by integrating the 
Ordinary Differential Equation (ODE) system (\ref{ODE}). 

In both cases, the procedure has been used to interpolate curves on Shape Space (\cite{Le2003}, \cite{Kume2007}). 
On the other hand, in \cite{Huckemann2010} the LC parallel transport has been used to translate a deformation from a shape to a different one. 

This use of the parallel transport is very interesting but,  in many fields, especially in elasticity, is not suitable because deformations pertain naturally to the Size-and-Shape Space. 
In fact, as rotations correspond simply to changes of observer, it is useful  to filter off it, while the change in size cannot be neglected,  being an important part of the elastic strain. So, in order to preserve the information about size changes, we will carry out all the parallel transports in the Size-and-Shape Space. Nevertheless the Levi Civita connection on the Size-and-Shape Space is known for both $m=2$ and $m=3$ (\cite{Kendall1999}), an explicit representation for the parallel transport along geodesics has never been given not even for $m=2$. In the latter case, by using the complex notation and following the approach of  \cite{Le2003} and \cite{Kume2007} adapted to the Size-and-Shape Space an explicit representation can be given (for details see \cite{Varano2014}). 
It is easy to show that,  in case of the Size-and-Shape Space the three hipotesys of the theorem 1 of Le (2003) become only two. In order the field $w(t)$ to be  parallel along a curve $\mu(t)$ must hold:
\begin{enumerate}\itemsep1pt
\item $w(t)$ must be horizontal i.e. Im$(\mu^*(t){w}(t))=0$.
\item $\dot w(t)$ must be vertical i.e. $\dot w(t)={\bf i}\lambda(t) \mu(t)$ where $\lambda(t)\in\mathbb{R}$.
\end{enumerate}
A geodesic in the Size-and-Shape Space has the form (\cite{Kendall1999}):
\begin{equation}\label{geod_SSa}
\mu(t)=\mu_0+t\,\mu_1 
\end{equation}
where Im$(\mu_0^{*}\mu_1)=0$.
By using \ref{geod_SSa} the two previous points became a differential algebraic system of equations that can be explicitly solved.
In particular, let $\mu_a$ and $\mu_b$ two aligned configurations. 
The  geodetic segment in the Size-and-Shape Space passing for the two points is:
\begin{equation}\label{geod_SSb}
\mu(t)=\mu_a + t (\mu_b - \mu_a)=\mu_0 + t \mu_1   
\end{equation}
where $t\in[0,1]$ is not the arc-length. 
Let $w_a$ be an horizontal vector at $\mu_a$, the parallel transported vector on $\mu_b$, will be:
\begin{equation}\label{PT_SS}
w_b=w_a-{\bf i} \frac{\text{Im}\left(\mu_b^{*}w_a\right) }{\left(\mu_a^{*}\mu_b+\|\mu_a\| \|\mu_b\|\right)\|\mu_b\|}\left(\|\mu_b\|\mu_a+\|\mu_a\|\mu_b\right)   
\end{equation}
where $\left(\mu_b^{*}w_a\right)$ is the complex inner product.

In the following sections, we shall compare the results obtained using two different transport techniques, 
the LC parallel transport in Size-and-Shape Space and our Direct Transport. 
%
Here, we can anticipate that the results obtained with the first technique are not suitable from our point of view, 
and we discuss some general issues concerning the concept of `same' deformation applied to different shapes. 

As the connection on the Size-and-Shape Space is inherited from that on the Configuration Space, 
we concentrate on the study of the latter.  The Levi Civita parallel transport in the Configuration Space corresponds simply to a linear shift of the coordinates, being $\Cc^k_m$ considered as equivalent to a 
$k\times m$-dimensional Euclidean space: $\Cc^k_m\equiv\Ec_{km}$. 

We want to stress a point, often implicitly assumed: the Kendall Shape-Space is non-Euclidean 
because of the removal of size and rotation, and not for some peculiar feature of the Configuration Space. 
This point is not obvious. 
Actually, the Configuration Space cannot be confused with the $m$-dimensional Euclidean Ambient-space. 
The question is: why the $k\times m$ configuration space should be considered Euclidean by itself?
In order to answer to this question it is necessary to understand the meaning of a vector $V_X$
belonging to the tangent bundle $\Tc \Cc^k_m|_X$ at the point $X$ of the Configuration Space . 

\begin{figure}
\begin{center}
\includegraphics[scale=0.6]{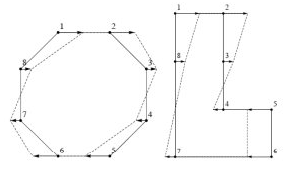} 
\end{center}
\caption{The Levi Civita parallel transport in the configuration space.}
\label{LC}
\end{figure}

A vector $V_X\in\Tc \Cc^k_m|_X$ may be represented as a $k\times m$ matrix, whose rows 
should be interpreted as the displacements of the $k$ landmarks, that is, each row is 
a vector of the Euclidean $m$-dimensional space, see Figure \ref{LC} left.  
Let us consider two configurations $Y$ and $X$; we may write 
\begin{equation}\label{LCtranspo}
Y=X+V_x\,
\quad\Leftrightarrow\quad
\left|\begin{array}{c} y_1 \\ \vdots \\ y_k \end{array}\right| =
\left|\begin{array}{c} x_1 \\ \vdots \\ x_k \end{array}\right| +
\left|\begin{array}{c} v_1 \\ \vdots \\ v_k \end{array}\right| \,,
\textrm{ with }
y_i\,,x_i \in \Ec_m\,, v_i \in \Vc\Ec_m\,.
\end{equation}
Transporting $V_X$ from $X$ to a different configuration $\bar X$ by using the Levi Civita connection 
consists in displacing the different landmarks of $\bar X$ with the same displacements applied to $X$, 
that is (see Figure \ref{LC} right):
\begin{equation}\label{LCtranspo2}
\bar Y=\bar X+V_{\bar x}\,
\quad\Leftrightarrow\quad
\left|\begin{array}{c} \bar y_1 \\ \vdots \\ \bar y_k \end{array}\right| =
\left|\begin{array}{c} \bar x_1 \\ \vdots \\ \bar x_k \end{array}\right| +
\left|\begin{array}{c} v_1 \\ \vdots \\ v_k \end{array}\right| \,,
\textrm{ with }
\bar y_i\,, \bar x_i \in \Ec_m\,, v_i \in \Vc\Ec_m\,.
\end{equation}
A first consequence of such a transport is that, in general, even affine deformations are not preserved.
Actually, affine deformations of $\Ec_m$ 
can be univocally represented by an element $H$ of the Lie Group $GL(m,\Real)$,
apart from a translation $\in\Vc\Ec^m$, so that $V_X=X\, H$;
thus, a meaningful transport of the vector $V_X$ to point $X$, which correspond to the same element $H$, 
is $V_{\bar X}= \bar X\,H$.
%
In the following, we shall propose an algorithm to define a connection on the configuration space which preserves deformations, at least for affine cases:
\begin{itemize}
\item Define a parallel transport $\tau$ which preserves deformations (see in the next (\ref{tau}));
\item Construct the connection $\nabla$ associated with $\tau$ (see (\ref{tau_to_nabla});
\item Endow the manifold with a new Riemannian metric compatible with $\nabla$. 
\end{itemize}
Note that we do not require the torsion of the connection to be null; 
as a consequence, we lose the uniqueness, and we cannot directly transfer the connection from the Configuration Space 
to the Size-and-shape Space, and eventually to the Shape Space; moreover,  
in passing from one space to another we will have to make some appropriate choices.
%
%
%
%
\begin{figure}
\begin{center}
\includegraphics[scale=0.6]{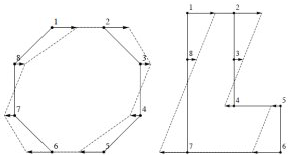}
\end{center}
\caption{An affine component preserving parallel transport in the configuration space.}
\label{DT}
\end{figure}
\subsection{The Direct Transport.}
%
%
\begin{figure}[t]
\begin{center}
\includegraphics[scale=0.2]{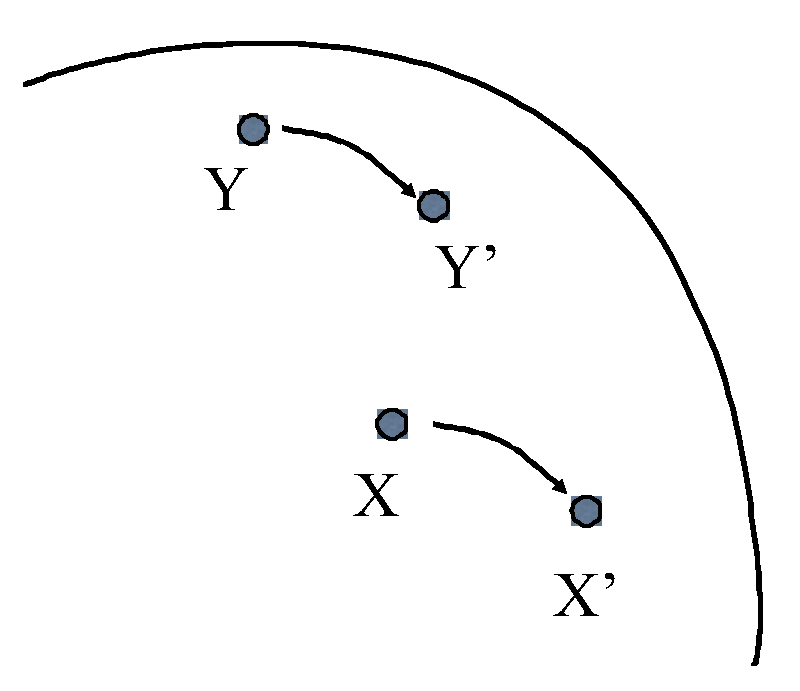}
\end{center}
\caption{Equipollence between point pairs.}
\label{gino}
\end{figure}
In order to define on the Configuration Space a parallel transport based on the deformation,
we follow the approach used in \cite{Schouten1954} to introduce the left invariant connection on Lie groups. 

Two pointpairs $(X, X')$ and $(Y, Y')$ of the Configuration Space are called \emph{equipollent} 
if the are related by the same diffeomorphisms $\Phi$ of $\Ec_m$, that is, if
$X'=\Phi(X)$ and $Y'=\Phi(Y)$, see Figure \ref{gino}.
Then, given a pointpair $(X, \Phi(X))$, it is possible to 
generate other equipollent pointpairs, that is, to transport  the pair $(X,\Phi(X))$ to $(Y,\Phi(Y))$.

If pairs are close enough, we obtain a transport of vectors, called Direct Transport (DT) 
because it does not depend on the path. This approach works well in the cases of $g$-spaces, in which a Lie group of transformations acts transitively on a manifold. 

In that case, the dimension of the Lie group equals the dimension of the manifold, and there is a one to one correspondence between each tangent space of the manifold and the Lie algebra of the group of transformations: 
each vector in the tangent space corresponds to one element of the Lie algebra of the transformations group. 

In our case, the notion of equipollence is meaningful but the group of the diffeomorphisms of $\Ec_m$ is 
infinite dimensional;  then, given a point pair  $(X,X')$ we have an equivalence class of infinite diffeomorphisms transforming $X$ in $X'$: we need a rule for selecting the one representing the point pair. 

This problem does not exist in the case of affine diffeomorphisms because, as previously observed, in that case the whole deformation can be represented by an element $H\in GL(m,\Real)$. For this reason we start by proposing a decomposition  of the tangent bundle $\Tc\Ec_m$ as the direct sum of a Uniform Component $\Uc\Tc\Ec_m$, 
and a Non Uniform Component $\Nc\Uc\Tc\Ec_m$; thus: $\Tc\Ec_m=\Uc\Tc\Ec_m \oplus^{\varphi} \Nc\Uc\Tc\Ec_m$. 
The first sub-bundle  is  the vector bundle defined, on each configuration $X$, as:
\begin{equation}
\Ac\Tc\Ec^m|_X=V_X \in \Tc\Ec^m : V_X= X H \hspace{2 cm} \forall H\in GL(m,\Real)
\end{equation}
The second subspace is defined as the complementary one with respect to the direct sum $\oplus^{\varphi} $ which will be defined in the following (\ref{g_phi}).    

%
\subsubsection{The Direct Transport of the Uniform Component in the Configuration Space}
%
Let us consider a pointpair of centered configurations \footnote{From now on, unless otherwise specified,
we will consider centered configurations: $X_C=X$.} 
$(X,X')$. 
We can easily calculate a linear approximation $F\in GL(m,\Real)$ of the map which transform $X$ in $X'$:
using the Penrose pseudo inverse $X^+=(X^T\,X)^{-1}X^T$, we have $F= X^+X'$. It is important to remark that
$X\,F=\bar X \ne X'$, unless $X$ and $X'$ are related by a linear transformation.
This allows to define the Direct-Transported pointpair on the configuration $Y$ as $(Y,Y')$, with: 
\begin{equation}
Y'=Y\, F= Y\,X^+ X'
\end{equation}
In order to pass from pointpairs to vectors we consider small deformations $F=I+hH$ where $h$ is a smallness parameter. Passing to the limit one has:
\begin{equation}
V_X=\lim_{h \to 0} \frac{X'-X}{h}=\lim_{h \to 0} \frac{X\,(I+h\,H)-X}{h}=X\,H
\end{equation}
\begin{equation}
V_Y=\lim_{h \to 0} \frac{Y'-Y}{h}=\lim_{h \to 0} \frac{Y\,(I+h\,H)-Y}{h}=Y\,H=Y\, X^+\,V_X
\end{equation}
Thus, we obtained  a parallel transport on the uniform component of the vector bundle:
	\begin{equation}\label{tau}
	\tau_{\Ac}: \Ac\Tc\Ec_m|_{\gamma(a)}\to\Ac\Tc\Ec_m|_{\gamma(t)}\hspace{2cm}V_a\mapsto V_t=\gamma(t)\,\gamma(a)^+\,V_a
	\end{equation}
It is easy to check that $\tau_t$ is a linear, one to one, map from  
$\Ac\Tc\Ec_m|_{\gamma(a)}$ to	$\Ac\Tc\Ec_m|_{\gamma(t)}$. 
This allows us to use equation (\ref{tau_to_nabla}) in order to obtain the coefficients of the connection. 
Here, it is important to remark that $\tau_t$ is path independent; 
then, the curvature of the connection vanish, and this gives rise to a notion of absolute parallelism. 
%
%
The last feature we need to define a Riemannian connection is the introduction of a Riemannian metric
on $\Ac\Tc\Ec_m|$, compatible with $\tau_{\Ac}$:
\begin{equation}\label{g_A}
g_{\Ac}(U_X,V_X)=g\left(X^+\,U_X, X^+\,V_X\right)=\text{trace}\left(U^T_X \left(X^{+T}\,X^+\right)\,V_X\right)
\end{equation}
In fact, a direct calculation shows that:
\begin{equation}
g_{\Ac}(U_X,V_X)=g_{\Ac}(U_Y,V_Y)\,,
\end{equation}
whenever $U_Y, V_Y$ are the directly transported of $U_X, V_X$ from $X$ to $Y$.

The metric $g_{\Ac}$, if  defined on the whole $\Tc\Ec_m|$ should be considered as a sub-Riemannian metric or singular Riemannian metric (\cite{Lee1997}).  
The integration of $g_{\Ac}$ along a segment of $\Ec_m|$ leads to a corresponding sub-Riemannian distance:
\begin{eqnarray}\label{affine_distance}
\nonumber d_{\Ac}(X,Y)&=& g_{\Ac}\left((Y-X),(Y-X)\right)= \\[3mm]
                      & & g\left(X^+\,(Y-X), X^+\,(Y-X)\right)=g\left( X^+\,Y-I_m, X^+\, Y-I_m\right)\,.
\end{eqnarray}
For affine transformations this distance is proportional to a \emph{mean elastic strain energy} 
gauged by the deformation $X^+\,Y-I_m$ relative to the configuration change from $X$ to $Y$ (\cite{Younes1998}). 
Let us note that the elastic energy stored in $\Bc$ under the deformation $\Phi$ is given by:
\begin{equation}
J_{\Ac}(\Phi)=\mu\int_{\Bc} (\nabla^{\Vc_m}\Phi-I_m)\cdot(\nabla^{\Vc_m}\Phi-I_m)\,,
\end{equation}
where $\nabla^{\Vc_m}$is the standard gradient in $\Vc_m$.
Incidentally, we note that, in the case of simplexes (triangles for $m=2$ and tetrahedra for $m=3$ ), 
when $\Real^{m(k-1)}=\Real^{m\times m}\equiv GL(m,\Real)$, the direct transport coincides 
with the left invariant connection on  $GL(m,\Real)$, 
and the metric $g_{\Ac}$ with the classical left invariant metric.

Finally, we remark that we have formulated two different notions of DT: 
the DT of pointpairs and the related Riemannian DT connection on the vector bundle $\Ac\Tc\Ec^m$. 
This has been done in order to show that the first is not simply a method, 
but it endows the Configuration Space with an additional geometrical structure. 
On the other hand, for the aim of this paper, i.e. the comparison between trajectories in the Shape space, 
the DT of pointpairs is directly applicable. For this reason in the following we will focus directly on that. 
\par
%
\subsubsection{The Direct Transport of the Uniform Component in the Size-and-Shape Space}
%
\par
Deformations pertain to the Configuration Space, but they also have sense in the Size-and-shape Space. 
In fact, as rotations correspond simply to changes of observer, in many fields, especially in mechanics, it is usual to substitute the concept of deformation with that of strain, by filtering off rotations; 
conversely, often the change in size is an important part of the strain. 

So, in order to preserve the information about size changes, 
we prefer to transport deformations in Size-and-shape Space
and to filter off size only after the data centering.
As previously said, without the LC connection, we don't have a direct way to transfer our DT connection in the quotient space, then we need to handle directly equivalence classes. A deformation from a size-and-shape 
to another must be defined as an equivalence class of transformations bringing from an equivalence class of 
configurations to another.

We refer to the equation (\ref{size-and-shape}) for the representation of the size-and shape $[X]_S$ of a configuration $X$. 
Let suppose to have a pointpair in the Size-and-shape Space, i.e., a pair of size-and-shapes $\left([X]_S,[X']_S\right)$. 
Given the linear part $F= X^+\,X'$ of the deformation from $X$ to $X'$, we define the equivalence class
of deformations $[F]_S$:
\begin{eqnarray}
\nonumber [F]_S &=&[X]_S^+ \,[X']_S = ([X]_S^T\, [X]_S)^{-1}\,[X]_S^T\, [X']_S\\[3mm]
                &=&\{ (Q^T\,X^T\,X\,Q)^{-1}\,Q^T\,X^T\, X'\,Q'\,:\,Q,Q'\in SO_m\}\}  \nonumber\\[3mm]
                &=&\{Q^T\,(X^T\,X)^{-1}\,X^T\, X'\,Q'\,:\,Q,Q'\in SO_m\}   \nonumber\\[3mm]
                &=& \{Q^T\,F\,Q':Q,Q'\in SO_m \}\,,
\end{eqnarray}
To strictly parametrize the size-and-shape change and obtaining the strain, the rotational component of $[F]_S$
has to be removed with a polar decomposition:
\begin{equation}\label{strain}
[V]_S=([F]_S[F]_S^T)^{1/2}=\{Q^T\,V\,Q:Q\in SO_m \}\,.
\end{equation}
We remark that the equivalence class $[V]_S$ always represents the same strain, only if one apply a certain element 
$Q^T\,V\,Q$ to the corresponding one $X\,Q$ of the shape class, sharing the same rotation $Q$; otherwise, 
one obtains a deformation with the same principal strains but different principal strain directions, 
and thus, a different shape. 

Now, even though the definition (\ref{strain}) of strain between a pair of size-and-shapes is unambiguous, the application of this strain to a  size-and-shape $[Y]_S$ different from $[X]_S$  becomes ambiguous:
it is not obvious which elements of $[V]_S$ and $[Y]_S$ should be selected. 
We are free to chose a conventional and reasonable rule: to  align $X$, $X'$ and $Y$, we can perform the DT 
passing directly through the configuration space. In practice, we formulate a suitable rule to select a triplet of \emph{optimally aligned} configurations $X^a\in[X]_S$, $X'^a\in[X']_S$, and $Y \in [Y]_S$, 
and define the directly transported of the pair $\left([X]_S,[X']_S\right)$ as 
the pair $\left([Y]_S,[Y']_S\right)$, where:
\begin{equation}\label{DT_size-and-shape}
[Y']_S=[Y\,(X^a)^+\, X'^a ]_S
\end{equation}
To choose the alignment rule, we consider an important distinction between alignments: the alignment within the 
pair $(X,X')$, and the alignment between the two undeformed configurations $X$ and $Y$. 
The first concerns two configurations related by an actual deformation, while the second is simply an alignment of shapes. This consideration justify the need of using two different notions of \emph{optimal alignment},
based on the Ordinary and the Modified Procrustes Analyses, (OPA) and (MOPA) respectively;
the difference will be explained in the next section.  

Here, we complete the definition of the DT (\ref{DT_size-and-shape}) 
by specifying that the alignment is implemented in the following order:
\begin{itemize}
\item $X$ is aligned with $Y$ using an OPA, yielding $X^a=X\,Q_{P}$
\item $X'$ is aligned with $X$ using a MOPA, yielding $X'^a=X'\,Q_{MP}$
\end{itemize}
where $Q_{P},\,Q_{MP}\in SO_m$ are the OPA and MOPA rotations, respectively.
Finally, the DT rule becomes:
\begin{equation}\label{DT_size-and-shape2}
[Y']_S=[Y\,(X\,Q_{P})^+\, X'\,Q_{MP} ]_S
\end{equation}
\par 
%
\subsubsection{The Modified-Ordinary and the Hierarchical Procrustes Analysis (MOPA \& HPA)}
%
\par
A common way to optimally align a configuration $Y$ with a configuration $X$ is to find the rotation minimising 
the size-and-shape distance $d_S(X,Y)$ as follows:
\begin{eqnarray}
\nonumber Q_{P}&=&\text{argmin} ||YQ-X||
                 =\text{argmin}\,\sqrt{\text{trace}\left( Y^T\,Y -2\,X^T\,Y\,Q +X^T\,X\right)}\\[3mm]
               &=&\text{argmax}\,\sqrt{\text{trace}(X^T\,Y\,Q)}\,.
\end{eqnarray}
It can be easily proved that the result is the transpose of the rotational component of the polar decomposition 
of the matrix $X^T\,Y$.

Here, we propose an alternative definition of an optimal alignment MOPA-like,
based on the notion of deformation. 
We replicate the previous procedure by using the sub-Riemannian distance $d_{\Ac}(X,Y)$ 
defined in equation (\ref{affine_distance}): we obtain:
\begin{eqnarray}\label{strain_distance}
\nonumber Q_{MP}&=&\text{argmin} ||X^+\,Y\,Q-I_m||=\\[3mm]
&&\text{argmin}\, \sqrt{\text{trace}\left(\, Y^T\, X^{+T}\,X^+\,Y -2\,X^+\,Y\,Q +I_m\,\right)}=\\[3mm]
&& \text{argmax}\, \sqrt{\text{trace}(X^{+}\,Y\,Q)}\,.
\end{eqnarray}
In this case, the optimal rotation comes from the polar decomposition of the tensor $F=X^{+}\,Y$, 
which represents the linear part of the transformation $X \mapsto Y$. Basing on this definition, aligning a shape with another means filtering rotations out from the affine part of the deformation; let us remark that
rotations, as defined in a standard procrustes alignment, are not deformation-based. 
Using a mechanical language, we can say that MOPA transforms a deformation in a strain.  

We call MGPA (Modified General Procrustes Analysis) an iterative loop of MOPAs intended to align a set of shapes 
by minimising (\ref{strain_distance}).
When tackling different sets of configurations, we define Hierarchical Procrustes Analysis (HPA)
the algorithm based on the following steps:
\begin{itemize}
\item Within each set choose a local reference $X_c$.
\item Perform a GPA among the local references and find the Grand Mean (GM).
\item Perform a loop of MOPA in each set to align the configurations with its proper $X_c$. 
\end{itemize}
$X_c$ can be the first of a sequence of configurations, or the mean of the entire sequence; 
in this last case, this local mean can be the local consensus after a preliminary local GPA. 
We remark that the aforementioned algorithm is prompted by our specific assumptions about the data set:
each set of configurations is supposed to be generated by a given actual body undergoing deformations; thus,
for the first alignment, that among the reference configurations of each set (pertaining to different bodies), we use a GPA;
for the second one, we use MOPA as it involve configurations of a same body.
\par 
%
\subsubsection{The generalization to the non-uniform component of the deformation}
%
\par
As previously mentioned, the notion of `same', that is, equipollent, deformation is unambiguous only in the uniform case. 
The extension of this notion to the non affine part of the deformation is not trivial. 
The group of diffeomorphisms of $\Ec_m$ is an infinite dimensional differentiable group; 
thus, it is not a Lie group, and it does not have the same dimension of $\Ec_m$: 
it is overdimensioned to represent a parallel transport on $\Ec_m$.  
More simply: for a given point pair $(X,X')$, there exist infinitely many diffeomorphisms transforming $X$ in $X'$. This formal problem has been extensively faced in the theory of metamorphoses and related papers (\cite{Trouve1995}, \cite{Miller2001}, \cite{Younes2010}, \cite{Trouve2011}). Here the aim is different and  we prefer to follow an approximate approach.
In order to reduce the dimension we need to chose a family of diffeomorphisms able to interpolate the deformation. 
We do not require this family to be a group, but we require the family to have the same dimension as $\Ec_m$. 
The most simple and used family of deformations, extensively used in geometric morphometrics, 
is given by the Thin Plate Spline (TPS) model. We choose TPS despite the problem that TPS transformations are not always diffeomorphisms (folds can appear), because the TPS has a close form representation.  For a completely diffeomorphic approach we could consider \cite{Cootes2008} or other alternatives.      
\begin{figure}[h]
\begin{center}
\includegraphics[scale=0.7]{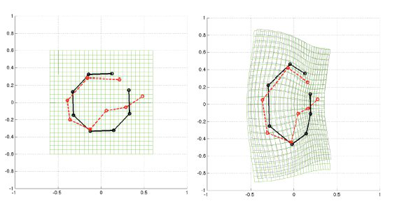}
\end{center}
\caption{TPS. Left: two undeformed configurations. Right: the same non affine transformation applied to both configurations each with its proper deformation grid in relation to its proper undeformed state on the left. The amount of non overlapping of the two grids indicates the inability to perfectly transport the entire deformation in the non affine case. }
\label{DT_TPS}
\end{figure}
\par\noindent
A  pair of thin plate splines is given by the bivariate function
\begin{equation}\label{TPS1}
y=\Phi(x)=(\Phi_1(t), \Phi_2(t))^T= c + Ax + W^T s(x),
\end{equation}
where $x$ is $(2 \times 1)$, $s(x)=(\sigma(x-x_1),..., \sigma(x-x_k))^T$, $(k \times 1)$ and
\begin{equation}
\sigma(t) = \left\{ \begin{array}{ll}
         ||h||^2 \log(||h||) & \mbox{if $ ||h|| > 0$};\\
        0 & \mbox{if $||h|| = 0$}.\end{array} \right.  
 \end{equation}
By using $k\times m$ matrices the equation (\ref{TPS2}) can be written as:
\begin{equation}\label{TPS2}
Y= l_k c^T + XA^T + S(X,X)W,
\end{equation}
where
\begin{equation}
S_{ij}(Y,X)=\sigma(y_i-x_j).
\end{equation}
The $2k + 6=2+4+2k$ parameters of the mapping are $c,A,W$.
There are 2k interpolation constraints in equation (\ref{TPS2}), and we introduce six more
constraints on $W$ in order to uncouple affine and non affine parts:
\begin{equation}\label{TPS_constraints}
 l_k^T W=0 \hspace{2 cm} X^T W=0
\end{equation}
The constrained interpolation problem \ref{TPS2} can be written as:
\begin{equation}
\begin{bmatrix}Y \\  0\\ 0 \end{bmatrix}=\begin{bmatrix}S(X,X) &  l_k & X \\[3mm]
   l_k^T & 0 & 0 \\ X^T & 0 & 0 \end{bmatrix}\begin{bmatrix}W \\  c^T\\ A^T \end{bmatrix}
\end{equation}
This problem, provided that $S(X,X)$ is invertible, has a unique solution, 
that has a closed form (\cite{Bookstein1989}, \cite{Dryden1998}): 

\begin{equation}
W = \Gamma^{11}\, Y\,,
\quad
\begin{bmatrix}c^T\\A^T\end{bmatrix} = \Gamma^{21}\, Y\,,
\end{equation}
with:
\begin{equation}\label{gamma}
\Gamma^{11} = S^{-1}(X,X)-S^{-1}(X,X)\,Q\,\Gamma^{21}\,,
\quad
\Gamma^{21}=(Q^T\,S^{-1}(X,X)\,Q)^{-1}Q^T\,S^{-1}(X,X)\,,
\end{equation}
and $Q=[l_k ,X]$.
It can be proved that the transformation of (\ref{TPS2}) minimizes the total bending energy of all possible interpolating functions mapping from $X$ to $Y$, where the total bending energy is given by:
\begin{equation}
J_{\Nc\Ac}(\Phi)=\int_{\Bc}  (\nabla^{\Vc_m}\nabla^{\Vc_m}\Phi\cdot\nabla^{\Vc_m}\nabla^{\Vc_m}\Phi)
\end{equation}
From a continuum mechanics point of view, this latter can be considered as a so-called \emph{mean second order strain energy}. The minimized total bending energy is given by:
\begin{equation}
J_{\Nc\Ac}(\Phi)=\text{trace}\left(Y^T\Gamma^{11}Y\right)
\end{equation}
where for $\Gamma^{11}$ hold the constraints: 
\begin{equation}\label{BE_constraints}
 l_k^T\, \Gamma^{11}=0 \hspace{2 cm} X^T\, \Gamma^{11}=0
\end{equation}
As known, once calculated $c,A,W$ we are able to draw the transformation grid representing the whole diffeomorphism in $\Ec^m$. 

How can we define a DT using these features? The concept is simple: we consider two point pairs $(X,X')$ and $(Y,Y')$ equipollent when they share (with a reasonable approximation) the same transformation grid. Given this definition the DT can be define as follows. 
With reference to Figure \ref{DT_TPS} let suppose to have the point pair $(X,X')$. We calculate the TPS parameter $c_X,A_X,W_X$ that \emph{interpole} the deformation between $X$ and $X'$ in such a way that:
\begin{equation}
X'= l_k c_X^T + XA_X^T + S(X,X)W_X
\end{equation}
We draw the associated transformation grid. On the undeformed grid we label the k landmarks corresponding to a different configuration $Y$ and we \emph{link} them to the grid.Then we apply the same diffeomorphism constraining the landmarks to follow the deformation of the grid.   
In practice we use this TPS parameters in order to \emph{extrapolate} the deformation toward the landmarks of the configuration $Y$. In this way the DT of the point pair $(X,X')$ from $X$ to $Y$ is defined as the new point pair  $(Y,Y')$ where:
\begin{equation}
\overline{Y}'= l_k c_X^T + YA_X^T + S(Y,X)W_X
\end{equation}
This definition of DT of point pairs is meaningful but has some limits: it corresponds to our definition of equipollence only approximatively:
\begin{itemize}
\item If one calculate the transformation grid on the new point pair  $(Y,Y')$ one finds a result slightly different from that calculated on $(X,X')$ (see Figure \ref{DT_TPS} )
\item If one perform a straight cycle, by transporting back $(Y,Y')$ on $X$ (on the same path) one finds a point pair slightly different from $(X,X')$
\end{itemize}
Now we would like to pass from DT of point pairs to DT of vectors.
First of all we note that, because we are working in centered configurations $X=CX$ and $X'=CX'$. As $C l_k c_X^T=0$ then:
\begin{equation}
X'=XA_X^T + CS(X,X)W_X
\end{equation}
and
\begin{equation}
Y'=C\overline{Y}'=YA_X^T + CS(Y,X)W_X
\end{equation}
Then, by using the constraint (\ref{gamma}) and (\ref{BE_constraints})
\begin{eqnarray}
X'-X&=&X\overline{\Gamma}_X^{21}X'-X + CS(X,X)\Gamma_X^{11}X'=\\
&=&X\left(\overline{\Gamma}_X^{21}X'-I_m\right)+CS(X,X)\Gamma_X^{11}(X'-X)
\end{eqnarray}
\begin{eqnarray}
\left(I_k-CS(X,X)\Gamma_X^{11}\right)(X'-X)=X\left(\overline{\Gamma}_X^{21}X'-I_m\right)
\end{eqnarray}
\begin{eqnarray}
\left(\overline{\Gamma}_X^{21}X'-I_m\right)=X^+\left(I_k-CS(X,X)\Gamma_X^{11}\right)(X'-X)
\end{eqnarray}
\begin{eqnarray}
\nonumber Y'-Y&=&Y\overline{\Gamma}_X^{21}X'-Y + CS(Y,X)\Gamma_X^{11}X'=\\
\nonumber&=&Y\left(\overline{\Gamma}_X^{21}X'-I_m\right)+CS(Y,X)\Gamma_X^{11}(X'-X)\\
\nonumber&=&YX^+\left(I_k-CS(X,X)\Gamma_X^{11}\right)(X'-X)+CS(Y,X)\Gamma_X^{11}(X'-X)\\
\nonumber&=&\left(YX^+\left(I_k-CS(X,X)\Gamma_X^{11}\right)+CS(Y,X)\Gamma_X^{11}\right)(X'-X)\\
&=&\Gamma_{XY}^3(X'-X)
\end{eqnarray}
Passing to the limit for $(X'-X)\to 0$ we obtain the DT rule for vectors:
\begin{equation}
V_Y=\Gamma_{XY}^3 V_X
\end{equation}
Then the proposed non affine DT on vectors is a linear map between vector spaces. Nevertheless it is easy to show that $\Gamma_{YX}^3\neq(\Gamma_{XY}^3)^{-1}$ then the DT cannot be reversed. Finally we can confirm that it is not really a Parallel Transport, than it does not lead to a formalized structure of connection. We can conclude that for the case of non affine deformations we have to consider the DT a \emph{method} rather than a formalized geometrical structure.
%

Nevertheless we can use a notion of distance in order to measure how much the non affine DT approximates a connection. 
A natural distance associated to the TPS is the bending energy (\cite{Bookstein1989}, \cite{Bookstein1997})
By using (\ref{BE_constraints}) we can write:
\begin{equation}
J_{\Nc\Ac}(\Phi)=\text{trace}\left((X')^T\Gamma^{11}X'\right)=\text{trace}\left((X'-X)^T\Gamma^{11}(X'-X)\right)
\end{equation}
Passing to the limit for $(X'-X)\to 0$ we can obtain a sub-Riemannian metric:
\begin{equation}\label{g_NA}
g_{\Nc\Ac}(U,V)=\text{trace}\left(U^T\Gamma^{11}V\right)
\end{equation}
Inasmuch as $\Gamma^{11}$ depends on $X$ then $g_{\Nc\Ac}(U,V)$ is different on each configuration. This sub-Riemannian metric can be considered as complementary with respect to $g_{\Ac}$ introduced for the affine part in equation (\ref{g_A}). In fact the first vanishes on affine deformations and the second is defined only on affine deformations.
This allows us to obtain a full Riemannian metric, which we can name an \emph{elastic metric} (\cite{Nardinocchi2014}), by combining the two singular metrics:
\begin{equation}\label{g_phi}
g_{\phi}=g_{\Ac}+g_{\Nc\Ac}
\end{equation}
By integration along a segment $(Y-X)$ we obtain the corresponding full order mean strain energy\footnote{The idea of endowing the shape space with two different metrics is of \cite{Bookstein1989}. On the other hand the bending energy is not a full-distance because: is singular and is not symmetric. The defined mean strain energy is not singular but is not symmetric. A standard notion of distance could be recovered by integrating $g_{\phi}$ along a geodesic with respect the LC connection induced by $g_{\phi}$ itself.}:
\begin{equation}
J_{tot}(\Phi)=\mu_1J_{\Ac}(\Phi)+\mu_2J_{\Nc\Ac}(\Phi)
\end{equation}
where $\mu_1,\mu_2$ are constitutive elastic coefficients.  
With respect to the elastic metric we can completely define the direct sum $\oplus^{\varphi}$ which allows the orthogonal splitting of the tangent bundle in $\Tc\Ec^m=\Uc\Tc\Ec^m \oplus^{\varphi} \Nc\Uc\Tc\Ec^m$.
This elastic metric calculated on transported deformations, can gauge how much the DT is similar to a  Riemannian parallel transport. The difference $g_{\phi}(U_X,V_X)-g_{\phi}(U_Y,V_Y)$ gives us information on how much $U_X,V_X$ are well transported on $Y$. 

%
\section{Case Study: Relating Deformations to PC scores}
%
Our purpose is to perform a reverse engineering experiment, i.e. to define a set of  deformations and to recover them via shape analysis. 

By means of this set of deformations we perform several experiments in order to show the effects of inter-subject variation and the capability of PCA in recovering these parameters. As stated in the Introduction, once scale, translation and rotation are removed from the set of configurations, 
ordination analyses are often used to find axes explaining decreasing amounts of morphological variation. The most used ordination technique is PCA performed on the covariance matrix. 
The meaning of PC axes is just that of “illustrating” a deformation. In the case of morphological trajectories, after a common GPA, we expect to visualize the deformations affecting single trajectories explained by using PCA. However, if an inter-subject difference between shapes  exists, the PCA will try to explain concomitantly 
both the intra- and inter- subject variation.

\begin{figure}[h]
\begin{center}
\includegraphics[scale=0.3]{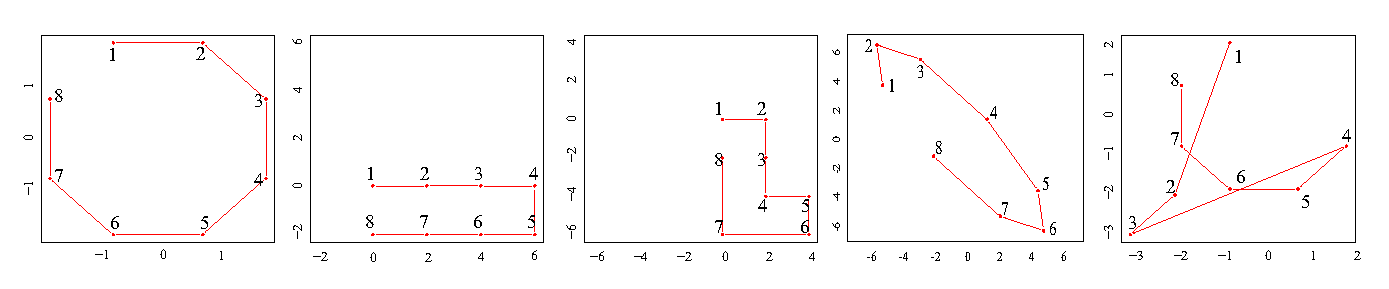}
\end{center}
\caption{The five reference bodies $\Bc_i$ used in this study.}
\label{Ref_Bodies}
\end{figure}

We discuss this issue by tackling an appropriate case study. 
Our study is posed in a 2D Euclidean space; 
to generate the dataset of morphological trajectories we consider a family of deformations 
$\Phi_c$ parametrised by a curve $t\mapsto c(t)=(\veps(t),\gamma(t))$,
and a set of five different reference bodies $\Bc_i$, see Fig. \ref{Ref_Bodies}, sampled with 8 landmarks, assumed homologous;
the two parameters $\veps$, $\gamma$ represent two different modes of deformation of the reference body.
In particular, we consider a uniform deformation, for which $\veps$ and $\gamma$ represent aspect ratio and shear,
and a non-uniform case, where $\gamma$ represents a bending curvature.
\begin{figure}[h]
\begin{center}
\includegraphics[scale=0.25]{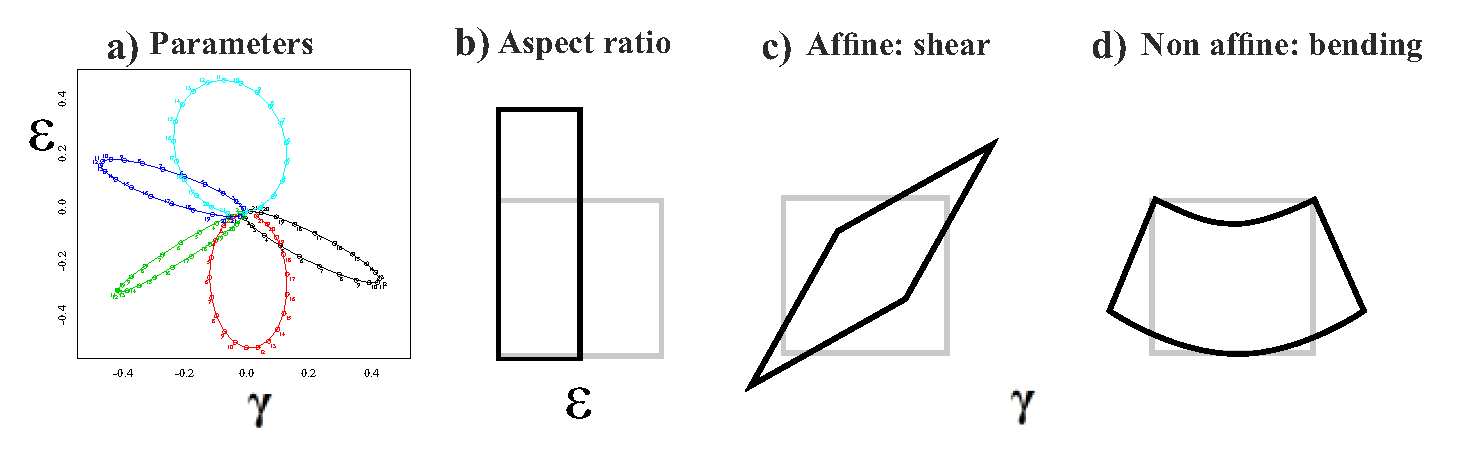}
\end{center}
\caption{Morphological meaning of deformation parameters cycles. a) The $\varepsilon-\gamma$ space of parameters. b) the morphological meaning of $\varepsilon$ parameter. c) the morphological meaning of $\gamma$ parameter in the affine case. d) the  morphological meaning of $\gamma$ parameter in the non affine case.}
\label{def_modes}
\end{figure}
\par
\begin{figure}[h]
\begin{center}
\includegraphics[scale=0.4]{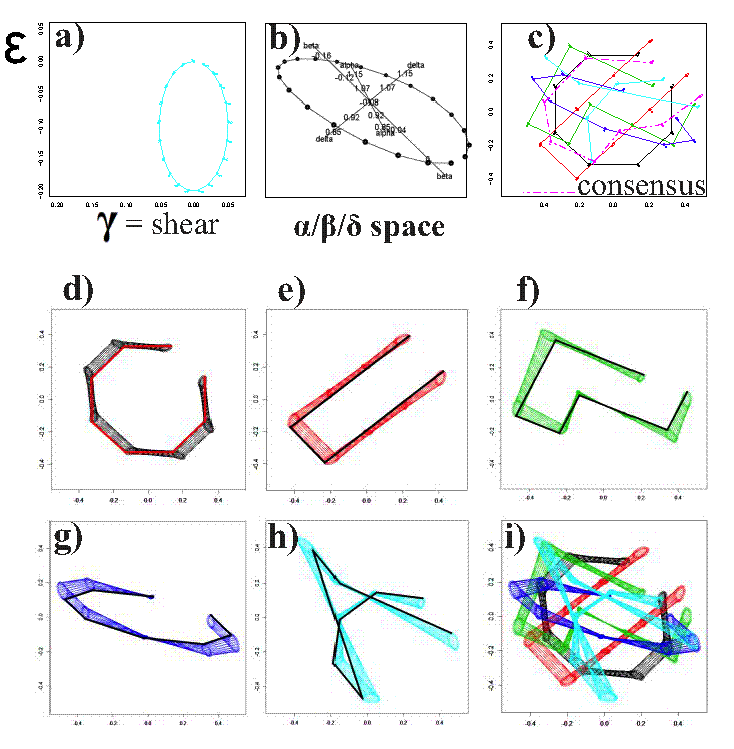}
\end{center}
\caption{ Case 1). The generation of the dataset with the same affine cycle applied to the five different shapes. a) parameters space. b) the parameter space identified by the symmetrical 2x2 $F$ matrix. c) preliminary optimally aligned initial shapes and their Grand Mean. d-i) deformed shapes. \label{dataset_1}}

\end{figure}

\begin{figure}[h]
\begin{center}
\includegraphics[scale=0.4]{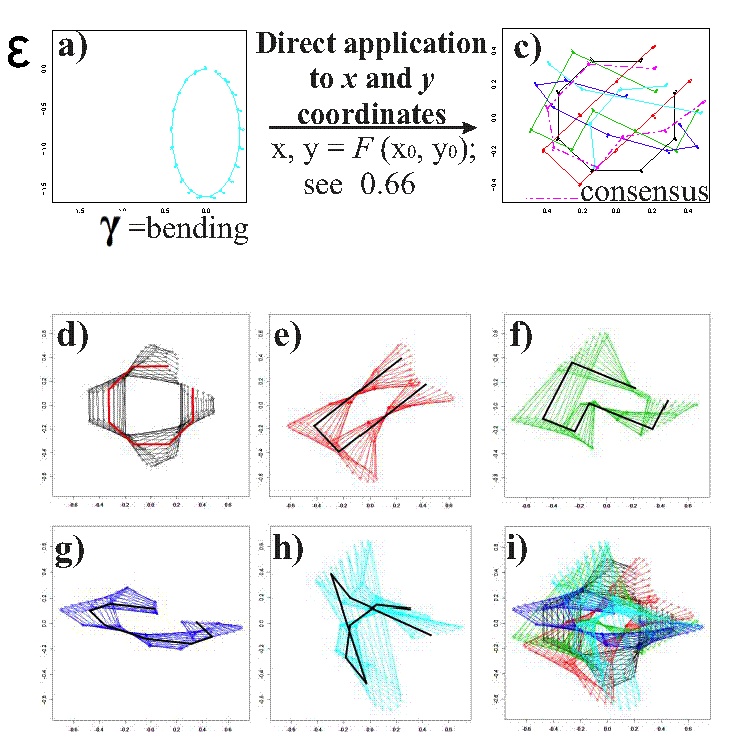}
\end{center}
\caption{Case 2). The generation of the dataset with the same non affine cycle applied to the five different shapes. a) parameters space. b) preliminary optimally aligned initial shapes and their Grand Mean. c-h) deformed shapes.}
\label{dataset_2}
\end{figure}

\begin{figure}[h]
\begin{center}
\includegraphics[scale=0.4]{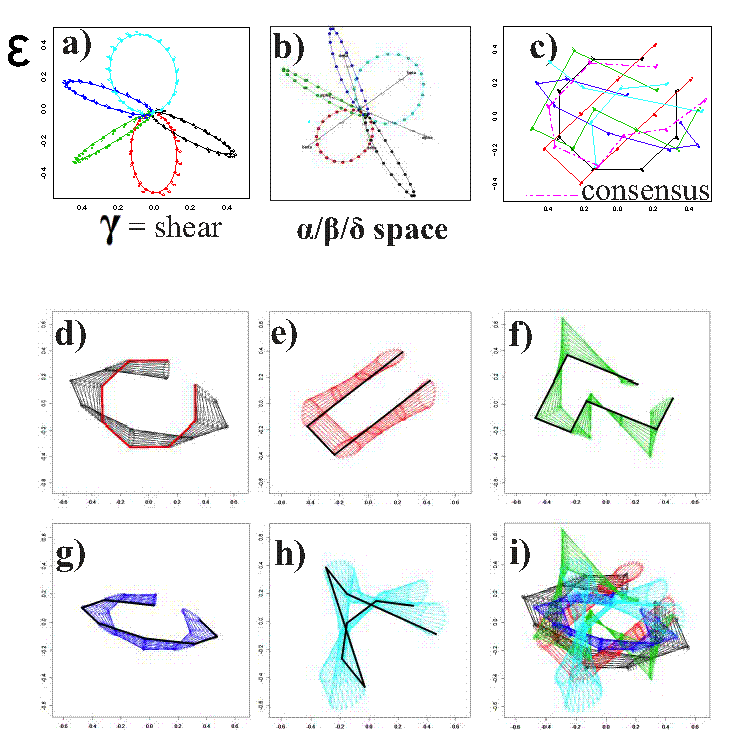}
\end{center}
\caption{Case 3). The generation of the dataset with different affine cycles applied to the five different shapes. a) parameters space. b) the parameters space identified by the symmetrical 2x2 $F$ matrix. c) preliminary optimally aligned initial shapes and their Grand Mean. d-i) deformed shapes}
\label{dataset_3}
\end{figure}

\begin{figure}[h]
\begin{center}
\includegraphics[scale=0.4]{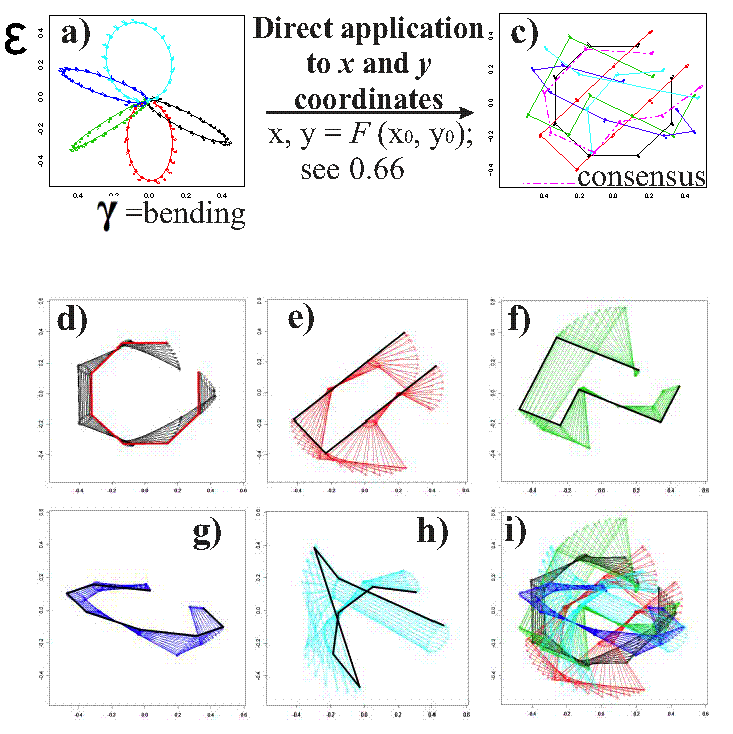}
\end{center}
\caption{Case 4). The generation of the dataset with different non affine cycles applied to the five different shapes. a) parameters space. b) preliminary optimally aligned initial shapes and their Grand Mean. c-h) deformed shapes.}
\label{dataset_4}
\end{figure}

\noindent
The uniform deformation is represented through a deformation matrix $F$ as follows
\begin{equation}\label{affine_def}
\begin{pmatrix} x \\[2mm]
                y \end{pmatrix}
                =
\begin{pmatrix} F_{11} & F_{12} \\[2mm]
                F_{21} & F_{22} \end{pmatrix}\,
\begin{pmatrix} x_o\\[2mm]
                y_o \end{pmatrix}\,,
                \quad\textrm{with }
                F=\exp\left[\begin{pmatrix} \varepsilon & 0 \\[2mm] 
                                             0          & -\varepsilon \end{pmatrix}
                           +\begin{pmatrix}  0 & \gamma   \\[2mm] 
                            \gamma & 0 \end{pmatrix}\right]\,,
\end{equation}
and $(x_o,y_o)\in\Bc_i$; 
let us note that $F$ maintains the area; note also that such a deformation is symmetric, and thus, 
it has a null rotational part.
The non-uniform deformation $\Phi_c$ is represented by
\begin{equation}\label{non_affine_def}
\begin{pmatrix} x\\[5mm]
                y\end{pmatrix}
   =\dfrac{1+\gamma\, \exp(\veps)\, x_o}{\gamma}\,\begin{pmatrix} 
              \sin\left(\dfrac{\gamma\, y_o}{\exp(\veps)}\right) \\[5mm]
              \cos\left(\dfrac{\gamma\, y_o}{\exp(\veps)}\right)-1 \end{pmatrix}\,.
\end{equation}
The morphological trajectories are generated by considering closed curves, called cycles, in the space of parameters;
we consider the following cases:
\begin{itemize}	
\item Case 1) One cycle $c(t)$ of uniform deformations as in (\ref{affine_def}) 
applied to the five reference bodies $\Bc_i$, see Fig.(\ref{dataset_1}):
\begin{equation}
\Tc_{F_c}(\Bc_j)=(\,X^j_{t_1}, \ldots X^j_{t_n}\,)
\quad\textrm{with $X^j_{t_i}$ configuration of $\Bc_j$ under $F_{c(t_i)}$};
\end{equation}
\item Case 2) One cycle $c(t)$ of non-uniform deformations as in eq. (\ref{non_affine_def}) applied to the five reference bodies $\Bc_i$, see Fig.(\ref{dataset_2}). :
$$
\Tc_{\Phi_c}(\Bc_j)=(\,X^j_{t_1}, \ldots X^j_{t_n}\,)
\quad\textrm{with $X^j_{t_i}$ configuration of $\Bc_j$ under $\Phi_{c(t_i)}$};
$$
\item Case 3) Five cycles $c^j(t)$ of uniform deformations as in (\ref{affine_def}) applied to the five reference bodies 
$\Bc_j$, that is, each different body $\Bc_j$ is deformed with a different cycle $c^j(t)$,
see Fig.(\ref{dataset_3}):
\begin{equation}
\Tc_{F_{c^j}}(\Bc_j)=(\,X^j_{t_1}, \ldots X^j_{t_n}\,)
\quad\textrm{with $X^j_{t_i}$ configuration of $\Bc_j$ under $F_{c^j(t_i)}$};
\end{equation}
\item Case 4) Five cycles $c^j(t)$ of non-uniform deformations as in (\ref{non_affine_def}) applied to the five reference bodies 
$\Bc_j$, that is, each different body $\Bc_j$ is deformed with a different cycle $c^j(t)$,
see Fig.(\ref{dataset_4}):
\begin{equation}
\Tc_{\Phi_{c^j}}(\Bc_j)=(\,X^j_{t_1}, \ldots X^j_{t_n}\,)
\quad\textrm{with $X^j_{t_i}$ configuration of $\Bc_j$ under $\Phi_{c^j(t_i)}$};
\end{equation}

\end{itemize}
In order to simulate a more realistic experiment, a random rotation is applied to each configuration of the generated datasets.
Let us note that the first two experiments are meant to assess if our procedure is able to 
recognise that the five trajectories have been generated by  
a same cycle of parameters $c(t)$, despite the differences among the reference bodies, and thus, 
among configurations along trajectories. 
We expect to obtain 	from the PCA a perfect agreement for case 1, and a good approximation for case 2. 

%

\subsection{Analysis of the datasets}
\begin{figure}[h]
\centerline{\includegraphics[scale=0.5]{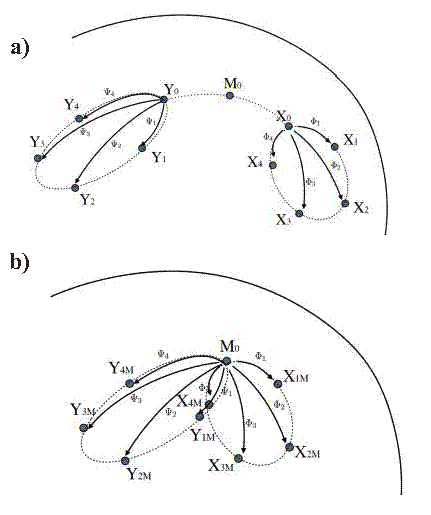}}
\caption{Data Centering via Direct Transport}
\label{data_centering}
\end{figure}

%
As previously explained our goal is to recover, for each analyzed case, the cycles of deformation parameters  via shape analysis. 
We compare three different methods:  classic GPA+PCA, data centering based on the Levi Civita parallel transport in Size-and-Shape Space and data centering based on our Direct Transport of point-pairs in the Size-and-Shape Space. Here we summarize the steps characterizing each one of the three methods (see Fig.(\ref{data_centering})):
\\
\vspace{0.5cm}

\par\noindent {\bf{Classic GPA+PCA.}}
\begin{enumerate}
\item Perform a GPA on the whole dataset and find the Grand Mean (GM).
\item Build the vectors $V^j_i=X^j_i-GM$.
\item Project $V^j_i$ on the tangent space to $GM$.
\item Perform a PCA in the tangent space to GM.
\end{enumerate}


\par\noindent {\bf{Data centering based on the Levi Civita connection in the Size and Shape Space.}} 
\begin{enumerate}
\item Hierarchical Procrustes Analysis
	\begin{enumerate}
	\item Within any of the $s$ cycles we choose a local reference $X^j_c$ that can be the first configuration of the cycle or the local mean (depending on the nature of the data: motion analysis or phenotypic trajectories analysis). In this paper we use the first configuration.
	\item Perform a GPA  with no scaling (in order to maintain the very nature of the deformation) between the $X^j_c$ and find the GM.
	\item We perform $s$ loops of OPA with no scaling to unit CS to align any shape of any cycle and its proper $X^j_c$. 
	\end{enumerate}
\item LC Parallel transport in the Size-and-Shape Space
	\begin{enumerate}
	\item Build the vectors $V^j_i=X^j_i-X^j_c$.
	\item Transport the vectors toward GM by using the explicit formula for the Levi Civita Parallel Transport in the Size-and-Shape Space (\ref{PT_SS}).
	\item add the transported vectors to the GM.
	\end{enumerate}
\item We then perform a standard GPA+PCA in the tangent space to GM.
\end{enumerate}

\par\noindent {\bf{Data centering based on the Direct Transport in the Size-and-Shape Space.}} 
\begin{enumerate}
\item Hierarchical Procrustes Analysis
	\begin{enumerate}
	\item Within any of the $s$ cycles we choose a local reference $X^j_c$ that can be the first configuration of the cycle or the local mean (depending on the nature of the data: motion analysis or phenotypic trajectories analysis). In this paper we use the first configuration.
	\item Perform a GPA  with no scaling (in order to maintain the very nature of the deformation) between the $X^j_c$ and find the GM.
	\item We perform $s$ loops of MOPA to align any shape of any cycle and its proper $X^j_c$. 
	\end{enumerate}
\item Direct Transport 
	\begin{enumerate}
	\item We perform a TPS analysis from any $X^j_c$ and the shapes within the cycles by  obtaining $s\times n$ TPS parameters $(c^j_i, A^j_i, W^j_i)$.
	\item We then apply the TPS transformations to the GM of the previous GPA. 
	\end{enumerate}
\item We then perform a standard GPA+PCA in the tangent space to GM.
\end{enumerate}
No reflections are allowed in our GPA (or MGPA) alignments.
%
\subsection{Results}
%
%
\begin{figure}[h]
\begin{center}
\includegraphics[scale=0.6]{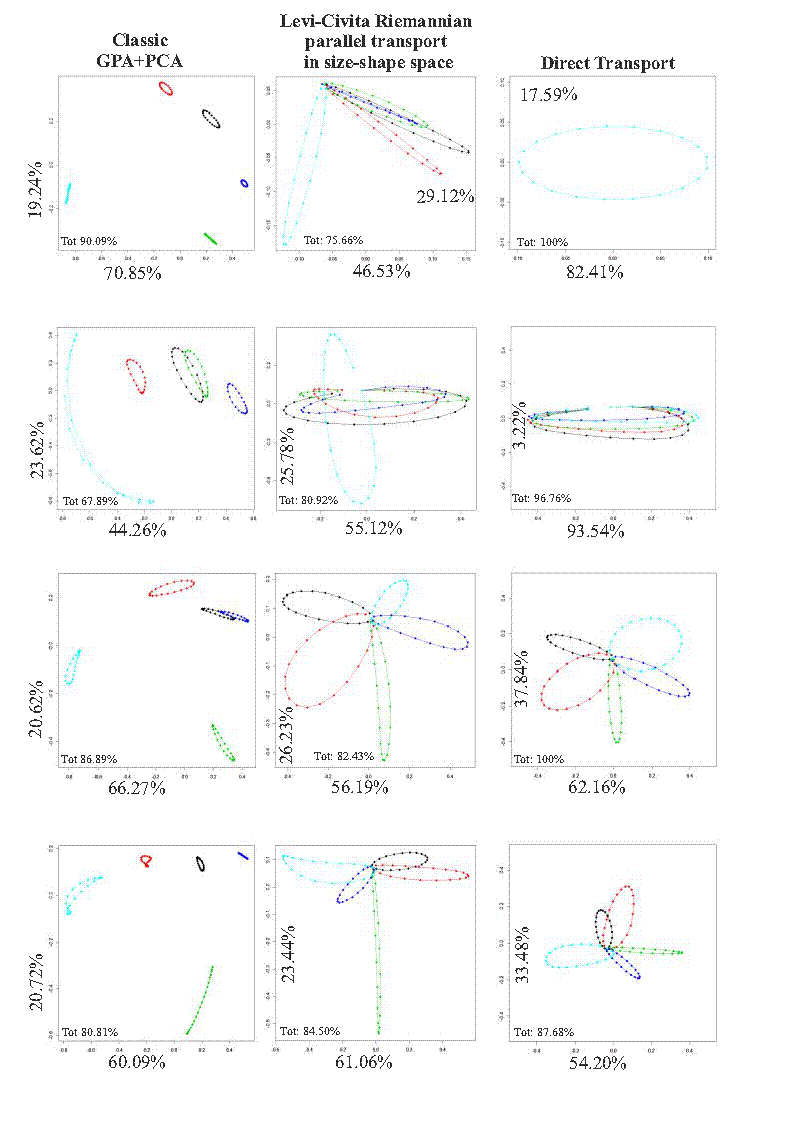}
\end{center}
\caption{The comparison of results of classic GPA+PCA, Levi Civita Riemannian Parallel Transport and Direct Transport. The first two PC scores are shown as well as their individual and global explained variances. From up to down the four above mentioned cases are presented by row, Case 1)-Case 4)}
\label{PCA}
\end{figure}
\begin{figure}[h]
\begin{center}
\includegraphics[scale=0.7]{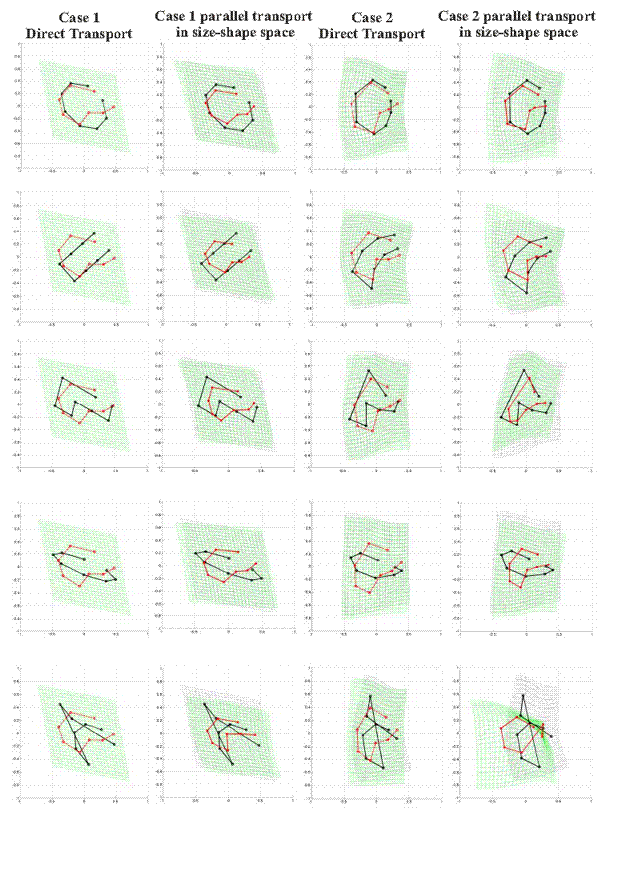}
\end{center}
\caption{The deformation grids between the first and the 10th  observation, i.e. that corresponding to the largest deformation within a cycle, for any shape for the first two cases.  In columns the Direct Transport and the Levi Civita Riemannian Parallel Transport are contrasted. Any graph presents two deformation grids superimposed: that corresponding to the real deformed shape (grid in grey, shape in black) and that corresponding to deformation transported on the Grand Mean (grid in green, shape in red). When these two grids coincide only one grid is visible. The degree of overlapping between these two grids is a visible measure of the goodness of the connection. Note that the Direct Transport perfectly recovers the deformation in the affine case and always works better than Levi Civita Connection in the non affine case.}
\label{grids_1_2}
\end{figure}
\begin{figure}[h]
\begin{center}
\includegraphics[scale=0.7]{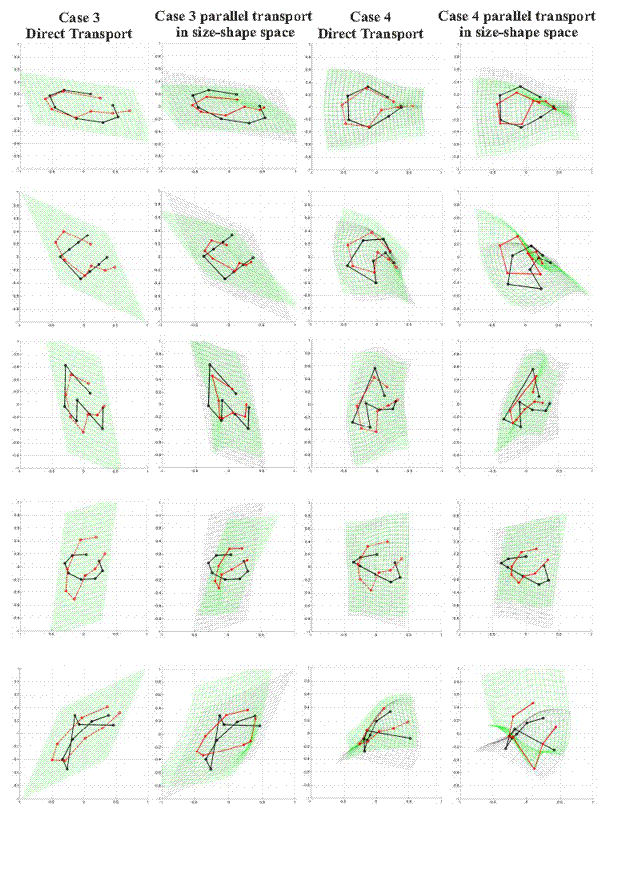}
\end{center}
\caption{The same representation of Fig. 12 for the third and fourth case.}
\label{grids_3_4}
\end{figure}
The comparison of results of classic GPA+PCA, Levi Civita Parallel Transport in the size and shape space and Direct Transport are plotted in Fig. (\ref{PCA}). The results relative to the Levi Civita  Parallel Transport in the shape space are shown in Supplementary Figure S1. There the first two PC scores are shown as well as their individual and global explained variances. From up to down the four cases and from left to right the three methods are reported.
The results clearly show that classic GPA+PCA is not a proper strategy to handle the problem of recovering deformation cycles. When applying a data centering, our Direct Transport perfectly recovers the deformation in the affine case, while it never happens for the Levi Civita connection. Moreover, in the non affine case the difference between the two approaches is clearly visible in both the PCA space and in the variance explained by each PC. We note that the variance explained can be used to evaluate the goodness of the deformation transport toward the Grand Mean. Another indication of the performance of the Direct Transport is given in Fig. (\ref{grids_1_2}) and (\ref{grids_3_4}) where we depicted the deformation grids between the first and the $10^{th}$ observation (i.e. that corresponding to the largest deformation) within each cycle for any shape for any case we analyzed above. We superimposed the deformation grids of the actual deformed shape and of the deformation transported toward the Grand Mean. When they coincide the original deformation is perfectly transported as it happens for the affine cases under the Direct Transport. In the non affine cases the grids non-overlapping under the Direct Transport is always visibly smaller than under the Levi Civita connection. We stress that the initial shapes we used here were intentionally challenging as they are hugely different. 
\subsection{Examples with real data: Left ventricle analysis}
\begin{figure}
\begin{center}
\includegraphics[scale=0.5]{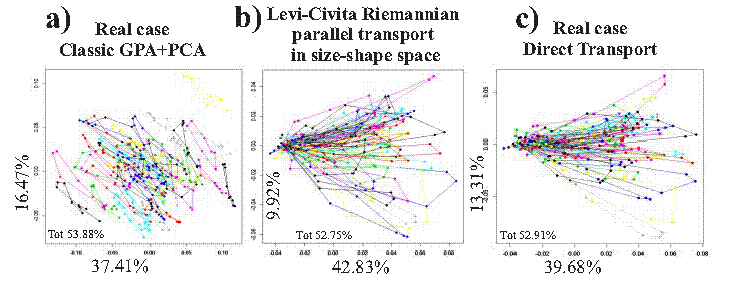}
\end{center}
\caption{Results for the real data.}
\label{PCA_hearts}
\end{figure}

In order to apply our procedures to real data, we used data coming from 3D echocardiography on 49 real human left ventricles (LV) moving in time and belonging to healthy subjects . These data come from the same research project partially published in Piras et al 2014. We collected shape data by means of 3D-STE (PST–25SX Artida, Toshiba Medical Systems Corp., Tokyo, Japan) . The final LV geometry is reconstructed by starting from a set of 6 homologous landmarks, manually detected by the operator for all subjects under study. The manual detection for a given set of landmarks is crucial because it allows recording spatial coordinates in perfectly comparable anatomical structures of different subjects (following a homology principle). In fact, completely automated approaches suffer from error of pattern identification depending on specific algorithms used for reconstruction. The result of our 3DSTE system is a time-sequence of configurations, each constituted by 1297 landmark, assumed to be homologous, for both the epicardial and endocardial surfaces, positioned along 36 horizontal circles, each comprised of 36 landmarks, plus the apex (Fig. \ref{PCA_hearts}a).
\begin{figure}
\begin{center}
\includegraphics[scale=0.6]{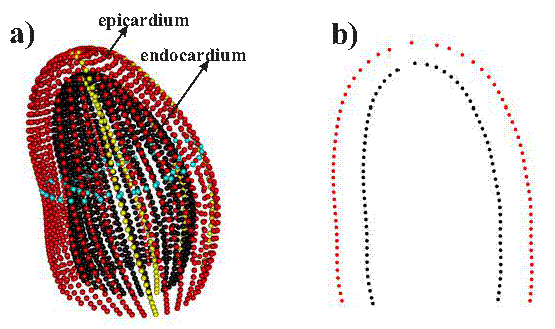}
\end{center}
\caption{Results for the real data.}
\label{data_hearts}
\end{figure}

It was possible to obtain the landmark cloud (upon which the standard rotational, torsional and strain parameters are computed and outputted by each Artida machine) by an unlocked version of the software equipping our PST–25SX Artida device, thanks to a special opportunity provided in the context of an official research and development agreement between the Dipartimento di Scienze Cardiovascolari, Respiratorie, Nefrologiche Anestesiologiche e Geriatriche, ‘‘Sapienza’’ Universita` di Roma and Toshiba Medical Systems Europe, Zoetermeer, The Netherlands. These 49 motion trajectories were acquired at the same electromechanically homologous times used in \cite{Piras2014}. Thus any individual trajectory consists of 9 time frames. In order to handle 2D data, we projected a coronal slice of the epicardial and endocardial landmarks cloud in 3D on the plane transversal to the LV base identified by the mitral annulus (Fig. \ref{data_hearts}). Fig. \ref{PCA_hearts}b reports the results of our procedures when applied to real data. While GPA+PCA still shows evidence of the ambiguous meaning of PCs that explain concomitantly intra- and inter- individual variation, the LC connection and the Direct Transport (both centered on the first time frame of the trajectories) yield similar results. We stress that the initial shapes we used in the simulated datasets were intentionally challenging as they are hugely different. In fact, the maximum geodesic Procrustes Distance between pairs of shapes in our simulated datasets is about 1.2 (the maximum allowed is 1.71=$\pi$/2). Our real data span 0.25 of geodesic Procrustes Distance thus making the use of LC Connection still acceptable. As these data represent healthy individuals, we speculate that future inclusions of pathological subjects could highly increase the sample variability (i.e the maximum geodesic Procrustes Distance) thus making the Direct Transport more efficient than LC connection. 

\subsection{Discussion and Future directions}
	In this paper we introduced the problem of comparing  morphological trajectories lying in very distant regions of the shape space. We proposed to solve this problem by performing a data centering in the Riemannian space, by means of a connection characterized by a parallel transport that preserves the deformations. In particular we pointed out that, in order to build  such a type of connection is not enough to give a metric and the related Levi Civita connection, but is necessary to introduce a connection with torsion. In the case of affine deformations we completely built a so-called Direct Transport connection, based on the left invariant connection in $GL(m)$. A set of simulations showed that this connection allows us to perfectly compare trajectories when only affine deformations are involved. Then we generalized the concept of Direct Transport   to the non affine case by proposing an approximated method for transporting vectors, based on the thin plate spline theory. Simulations showed that this method works very well. Nevertheless, when performed in the non affine case,  this method cannot be considered as a parallel transport because it is not reversible. Real data of human LV, suggest that when configurations differ for small Procrustes Distances the LC connection returns results very similar to the direct transport. Piras et al., 2014 used a linear shift procedure to center data in three dimensions. The more tuned approaches of LC connection or Direct Transport could lead to a more precise data centering of motion trajectories if the aim of a study is analyzing pure deformations.

\end{document}